%% file: main.tex
\documentclass{article}

\input{packages}

\newcommand{\ModelName}[0]{Foundation-Sec}

\newcommand{\llamathree}{Llama 3.1}
\newcommand{\deepspeed}{DeepSpeed}
\newcommand{\whiterabbit}{WhiteRabbitNeo-V2}
\newcommand{\chatgpt}{ChatGPT}
\newcommand{\gptthreefive}{GPT-3.5}
\newcommand{\gptfour}{GPT-4}
\newcommand{\gptfourOmini}{GPT-4o-mini}
\newcommand{\gptfourO}{GPT-4o}

\newcommand{\geminiflash}{Gemini Flash}
\newcommand{\securityllm}{SecurityLLM}
\newcommand{\primus}{Primus}
\newcommand{\secgemini}{SecGemini}

\newcommand{\ctibench}{CTIBench}

\newcommand{\secbench}{SecBench}
\newcommand{\cybermetric}{CyberMetric}

\newcommand{\alert}[1]{#1}

\title{Llama-3.1-FoundationAI-SecurityLLM-Base-8B \\ Technical Report}

\author{
    Paul Kassianik$^{1}$\thanks{Correspondence to: Paul Kassianik (\href{mailto:paulkass@cisco.com}{\texttt{paulkass@cisco.com}}) and Dhruv Kedia (\href{mailto:dkedia@cisco.com}{\texttt{dkedia@cisco.com}}).}, 
    Baturay Saglam$^{1,2}$, 
    Alexander Chen$^{1}$, 
    Blaine Nelson$^{1}$,
    Anu Vellore$^{1}$, 
    Massimo Aufiero$^{1}$,
    Fraser Burch$^{1}$, 
    Dhruv Kedia$^{1}$,
    Avi Zohary$^{1}$,
    Sajana Weerawardhena$^{1}$,
    Aman Priyanshu$^{1}$,
    Adam Swanda$^{1}$,
    Amy Chang$^{1}$,
    Hyrum Anderson$^{1}$,
    Kojin Oshiba$^{1}$,
    Omar Santos$^{3}$,
    Yaron Singer$^{1}$,
    Amin Karbasi$^{1}$\\
    $^{1}$Foundation AI -- Cisco Systems Inc.\\
    $^{2}$Yale University\\
    $^{3}$Security \& Trust Organization -- Cisco Systems Inc. \\
}

\date{April 2025}

\begin{document}
\maketitle

\begin{abstract}

    As transformer-based large language models (LLMs) increasingly permeate society, they have revolutionized domains such as software engineering, creative writing, and digital arts. 
    However, their adoption in cybersecurity remains limited due to challenges like scarcity of specialized training data and complexity of representing cybersecurity-specific knowledge. 
    To address these gaps, we present Foundation-Sec-8B, a cybersecurity-focused LLM built on the \llamathree{} architecture and enhanced through continued pretraining on a carefully curated cybersecurity corpus. 
    We evaluate Foundation-Sec-8B across both established and new cybersecurity benchmarks, showing that it matches \llamathree{}-70B and \gptfourOmini{} in certain cybersecurity-specific tasks. 
    By releasing our model to the public, we aim to accelerate progress and adoption of AI-driven tools in both public and private cybersecurity contexts.

\end{abstract}

\section{Introduction}

Artificial intelligence tools have rapidly become essential for boosting productivity and automating tasks across various domains. Frontier large language models (LLMs) such as \chatgpt{} \citep{gpt_4} have accelerated this trend, enabling users to complete in minutes tasks that once took hours or days. Since the release of \chatgpt{}, AI capabilities have advanced significantly \citep{Brown2020-nj, Wei2022-au, Schick2023-sr}. The rise of highly capable open-source LLMs \citep{llama2, llama3} has further democratized access, allowing researchers to adapt these models through fine-tuning and continued pretraining \citep{Gema2023-bp, Li2023-iv}. These adaptations often yield models that match or exceed proprietary counterparts on specialized tasks \citep{meditron, Wu2022-yo}.

Despite these advancements, the integration of LLMs into standard cybersecurity practices remains limited. 
Cybersecurity professionals face several barriers to adopting frontier LLMs, including restrictive safety guardrails from commercial providers, a lack of clean and publicly available cybersecurity datasets \citep{mbona2024datasets, Ammara2024-jn}, model hallucinations \citep{Tonmoy2024-kz, Ji2023-ev}, and challenges from distribution shifts \citep{Shaer2024-jm, Gupta2025-yg}. 
Meanwhile, serious threat actors have the compute and data resources to train custom, closed-source LLMs for nefarious purposes.
The broad and heterogeneous nature of cybersecurity -- from phishing detection to cryptographic analysis—makes developing a general-purpose security AI particularly difficult. As a result, the field is dominated by fragmented, task-specific tools \citep{Ferrag2024-wb}.

In this work, we introduce \textbf{\ModelName{}-8B}, a cybersecurity-specialized LLM built on \llamathree-8B \citep{llama3}. Our model shows significant performance gains over \llamathree{}-8B across cybersecurity benchmarks. 
We also release trained checkpoints to the research community, supported by comprehensive evaluations that highlight the model’s capabilities\footnote{\url{https://huggingface.co/fdtn-ai/Foundation-Sec-8B}}.
With this contribution, we advance research on applying LLMs to cybersecurity tasks and level the playing field for defensive cybersecurity practitioners.

\begin{figure}[tb]
  \centering
  \begin{subfigure}[b]{0.49\textwidth}
    \centering
    \includegraphics[width=\textwidth]{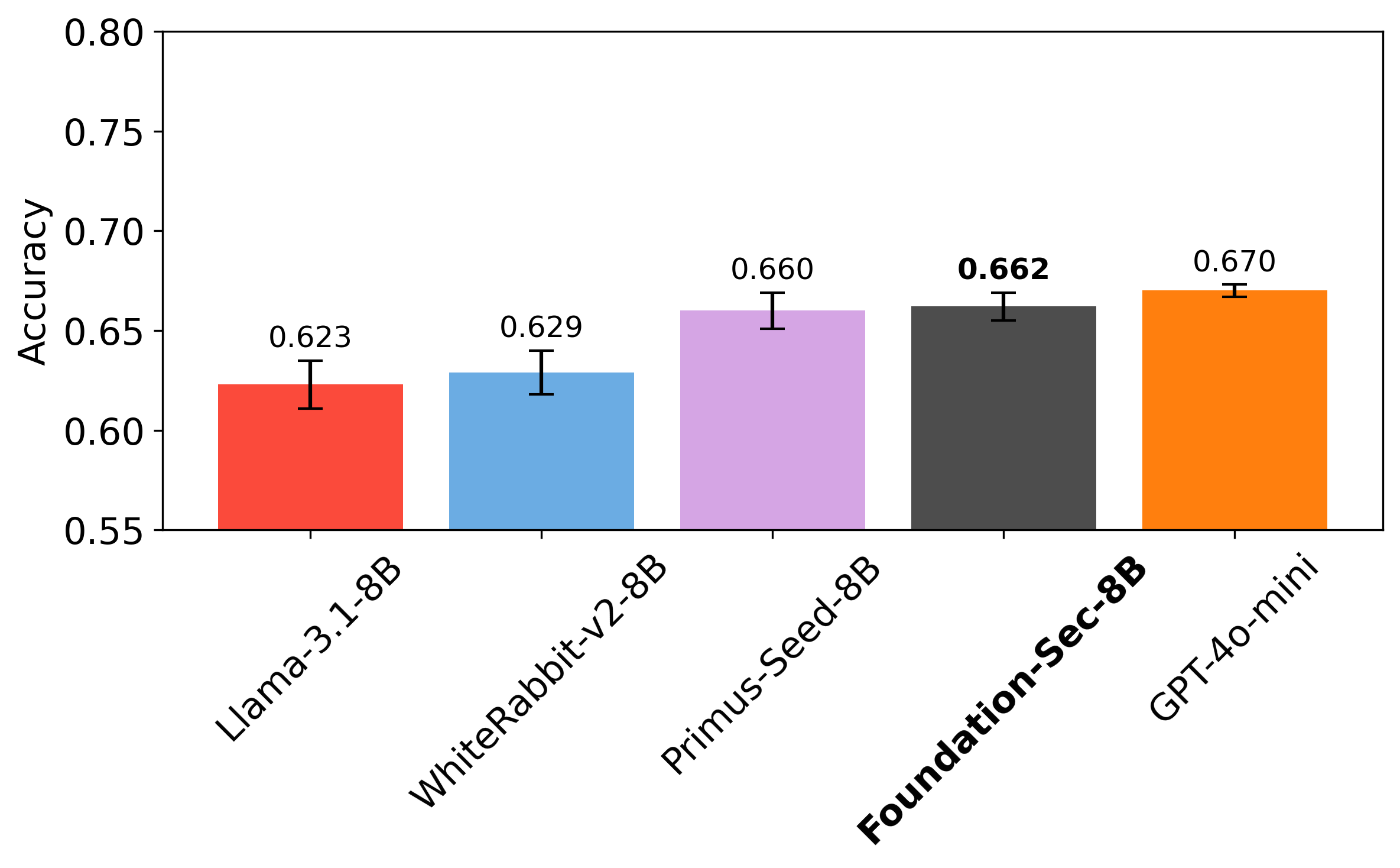}
    \caption{\ctibench-MCQA}
  \end{subfigure}
  \begin{subfigure}[b]{0.49\textwidth}
    \centering
    \includegraphics[width=\textwidth]{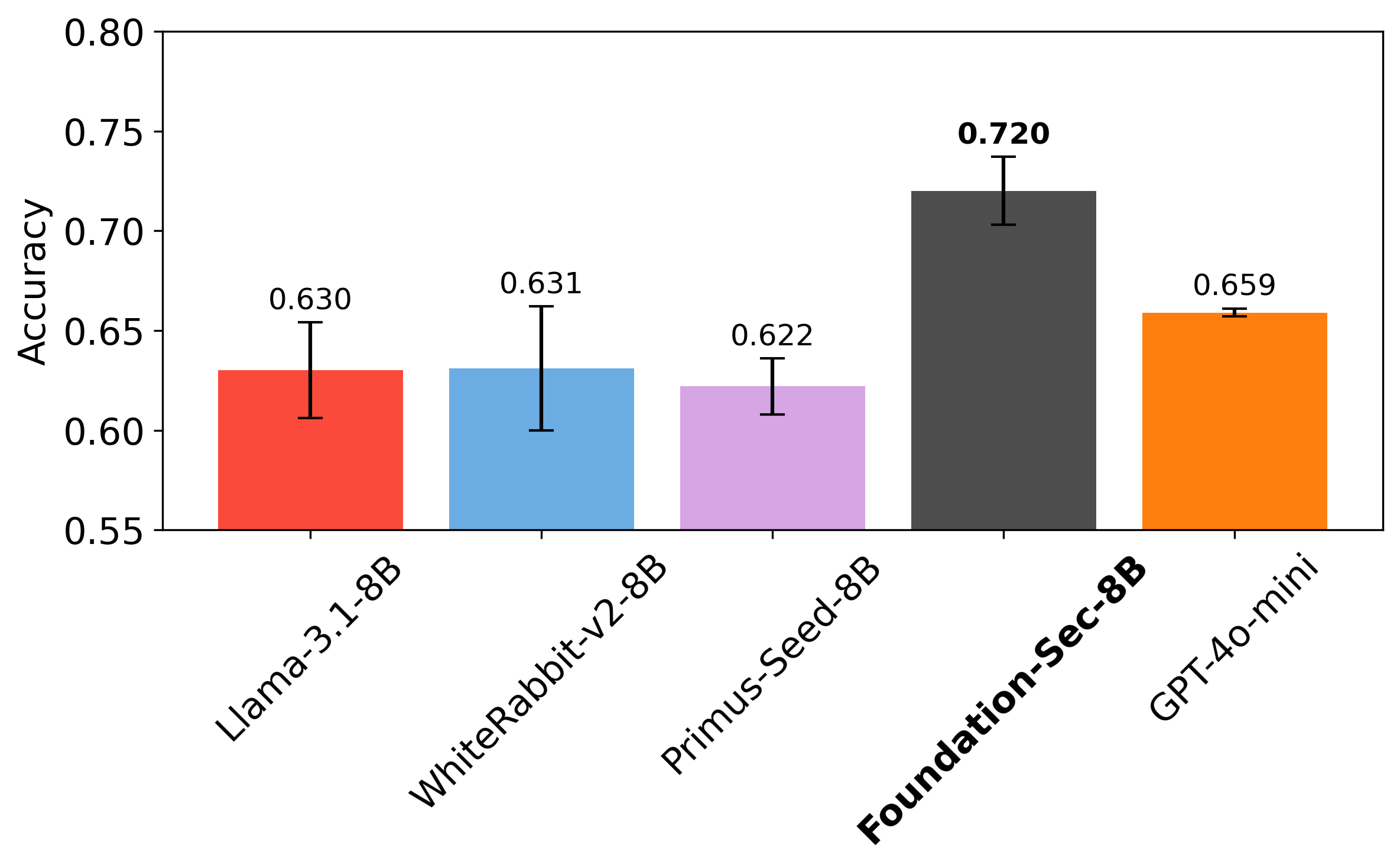}
    \caption{\ctibench-RCM}
  \end{subfigure}
 \caption{\textbf{Overview of core results on the selected cybersecurity benchmarks.} \ModelName{}-8B shows significant improvement over \llamathree{}-8B while matching or surpassing \gptfourOmini{} in cyber threat intelligence knowledge.}
  \label{fig:general-performance}
\end{figure}

\section{Related Works}

\subsection{Continued Pretraining}
Continued pretraining was initially proposed for transformer-based models as a way to improve domain-specific adaptation \citep{Hoang2019-hd, Wu2022-yo, Chakrabarty2019-tr, Gu2022-ro, Lee2020-ct, Gururangan2020-ks, Parmar2024-fq}. \citet{Wu2022-yo} showed that it enhances zero-shot and few-shot promptability, while \citet{Gururangan2020-ks} demonstrated that task-specific pretraining, even on unlabeled data, can boost model performance.

This approach has also been extended to LLMs, where models are continuously pretrained on billions of tokens to improve performance in specific domains such as medicine \citep{meditron, yuan2024continued}, law \citep{Colombo2024-ya, Colombo2024-ah}, mathematics \citep{azerbayev2023llemma, shao2024deepseekmath}, and code \citep{roziere2023code}. \citet{Parmar2024-fq, Ibrahim2024-do} further provide guidelines on when domain-specific continued pretraining is most effective.

\subsection{Benchmarks}

We reviewed a range of benchmarks and evaluation techniques at the intersection of LLMs and cybersecurity. Since our primary objective is to assess the knowledge of pretrained models, we exclude tasks that emphasize behavioral robustness over factual understanding.  

We have selected \alert{three} cybersecurity benchmarks framed as multi-class classification tasks. 
This design choice was motivated by the observation that pretrained models often fail to reliably follow instructions, rendering open-ended formats such as short answer questions (SAQ) unsuitable. 
We therefore focus on two formats: \textit{multiple choice question answering} (MCQA) and \textit{root cause mapping} (RCM).
Each MCQA question includes four options with a single correct ground truth answer.
For the RCM task, each Common Weakness Enumeration (CWE) \citep{cwe} description maps to exactly one CWE ID.

\subsubsection{Security Benchmarks}
\label{sec:benchmarks}

\paragraph{\ctibench{}}
Cyber threat intelligence (CTI) plays a key role in understanding and mitigating evolving cyber threats. 
\ctibench{} \citep{ctibench} targets practical, CTI-specific tasks and consists of five sections. 
However, most require advanced, consistent instruction-following capabilities and are not suitable for evaluation with a pretrained model. 
Therefore, we adapt two of them for base model evaluation: MCQA and RCM tasks.
MCQA includes 2,500 questions drawn from CTI frameworks such as NIST \citep{nist}, the Diamond Model of Intrusion Detection \citep{diamond_intrusion_detection}; regulations like GDPR \citep{gdpr}; CTI sharing standards such as STIX and TAXII \citep{stix_and_taxii}; and taxonomies like the MITRE ATT\&CK Framework \citep{mitre_attack} and CAPEC \citep{capec}. 
The RCM section evaluates a model's ability to identify the root cause of a vulnerability by linking CVE (Common Vulnerability Enumeration) \citep{cve} records and bug reports to CWE (Common Weakness Enumeration) entries \citep{cwe}. 
This task is highly nuanced, requiring a deep understanding of both CVE descriptions and the CWE taxonomy. 

\paragraph{\cybermetric{}}
The \cybermetric{} dataset \citep{cybermetric} was built from a collection of documents—such as NIST standards, research papers, public books, RFCs, and other cybersecurity publications—and converted into MCQA format using \gptthreefive{} with Retrieval-Augmented Generation (RAG). 
Human experts spent over 200 hours validating the questions and answers to ensure accuracy, relevance, and topic alignment. 
\cybermetric{} is available in four sizes: 80, 500, 2000, and 10,000 samples. 
We use the 500-sample version, as the larger sets contain only \chatgpt-generated questions on similar topics, without additional human validation.

\paragraph{\secbench{}}
\secbench{} \citep{secbench} includes both MCQA and SAQ questions, designed to assess LLM knowledge across two dimensions: Knowledge Retention and Logical Reasoning. A key strength of the benchmark is that a large portion of its content was created by human experts through a Cybersecurity Question Design Contest, making it more challenging than benchmarks like MMLU and \cybermetric{}. It is also the most recent in our collection, released in December 2024. Built using both contest submissions and open-source resources, \secbench{} includes English and Chinese sections. Since we are developing a monolingual model, we use only the English portion which contains 600 samples.

\subsubsection{Non-Security Benchmarks}

\paragraph{MMLU}
To evaluate whether continued pretraining on cybersecurity data affects the model’s general knowledge, we evaluate our model on the full Measuring Massive Multitask Language Understanding (MMLU) benchmark \citep{mmlu}.
MMLU covers a broad range of topics, including STEM, humanities, and social sciences, and serves as a strong indicator of a model’s retained knowledge across diverse domains. 
This allows us to check for any signs of overfitting or catastrophic forgetting, ensuring that domain specialization does not come at the cost of severe degradations to general reasoning and factual recall. 

\subsection{LLMs for Security}
\label{sec:llms_for_security}

Several works \citep{xu2024large, zhang2025llms} have reviewed the use of LLMs in cybersecurity. For a broader overview, we refer the reader to those studies and focus here on prior efforts most relevant to our work. While much research has focused on secure code generation \citep{silva2023repairllama, pearce2023examining, Zhang2023-ax}, our model is not designed or optimized for coding tasks. Instead, our primary goal was to integrate core cybersecurity knowledge into a pretrained language model, making it a strong foundation for downstream use cases (see Section \ref{sec:use_cases}). This enables broader applicability across cybersecurity tasks beyond code generation.

Prior to our work, several notable models were trained on general cybersecurity data. We include them as baselines in our evaluations, with the exception of the final model described below.

\paragraph{\whiterabbit} 
\citet{whiterabbitneo} introduced one of the earliest open-source cybersecurity LLMs, based on the \llamathree{} family. 
% It was released in 8B and 70B variants, focused on building uncensored models tuned for offensive security tasks.
Released in 8B and 70B variants, it focused on building uncensored models tuned for offensive security tasks.

\paragraph{\primus}
\citet{primus} curated a cybersecurity corpus from sources like MITRE, Wikipedia, cybersecurity companies, and manually collected cyber threat intelligence. Their dataset, containing nearly 2 billion tokens, was used to pretrain and instruct-finetune models based on \llamathree{}-8B.

\paragraph{SecurityLLM}
SecurityLLM\footnote{\url{https://huggingface.co/ZySec-AI/SecurityLLM}}, though not described in a technical report, is an open-source model based on Hugging Face’s Zephyr series \citep{zephyr} and built on Mistral 7B \citep{mistral_7b}. 
It was designed as a helpful security-focused assistant and finetuned across 30 domains, including attack surface threats, cloud security, and compliance frameworks like CIS Controls.

\paragraph{\secgemini}
\citet{secgemini} introduced the only frontier LLM specifically tailored for cybersecurity. \secgemini{} is a reasoning-enhanced \citep{openai_system_card_2024, snell2024scalingllmtesttimecompute} variant of \geminiflash{} \citep{google_gemini_flashlite}, augmented with live threat intelligence access with a knowledge-based system. 
As of this writing, it remains in closed preview and is not publicly available.

\section{Data Collection and Preparation}

One of the key challenges in developing LLMs for cybersecurity is the need for large volumes of high-quality data to drive meaningful downstream improvements. Further pretraining in other domains often relies on tens or hundreds of billions of tokens to effectively infuse domain-specific knowledge \citep{Yang_Band_Li_Candès_Hashimoto_2024}. In contrast, current cybersecurity models are typically trained on fewer than 2 billion tokens \citep{whiterabbitneo, primus}. Since cybersecurity is a relatively niche topic in the broader landscape of internet content, we developed a two-pronged approach to ensure diverse and high-quality data collection. First, we deployed general-purpose scrapers with a relevancy filter for broad data coverage. Second, we built custom scrapers targeting known high-quality cybersecurity sources and sites that are less accessible to URL-based scrapers.

\begin{figure}[h!]
\centering
\includegraphics[width=0.7\textwidth]{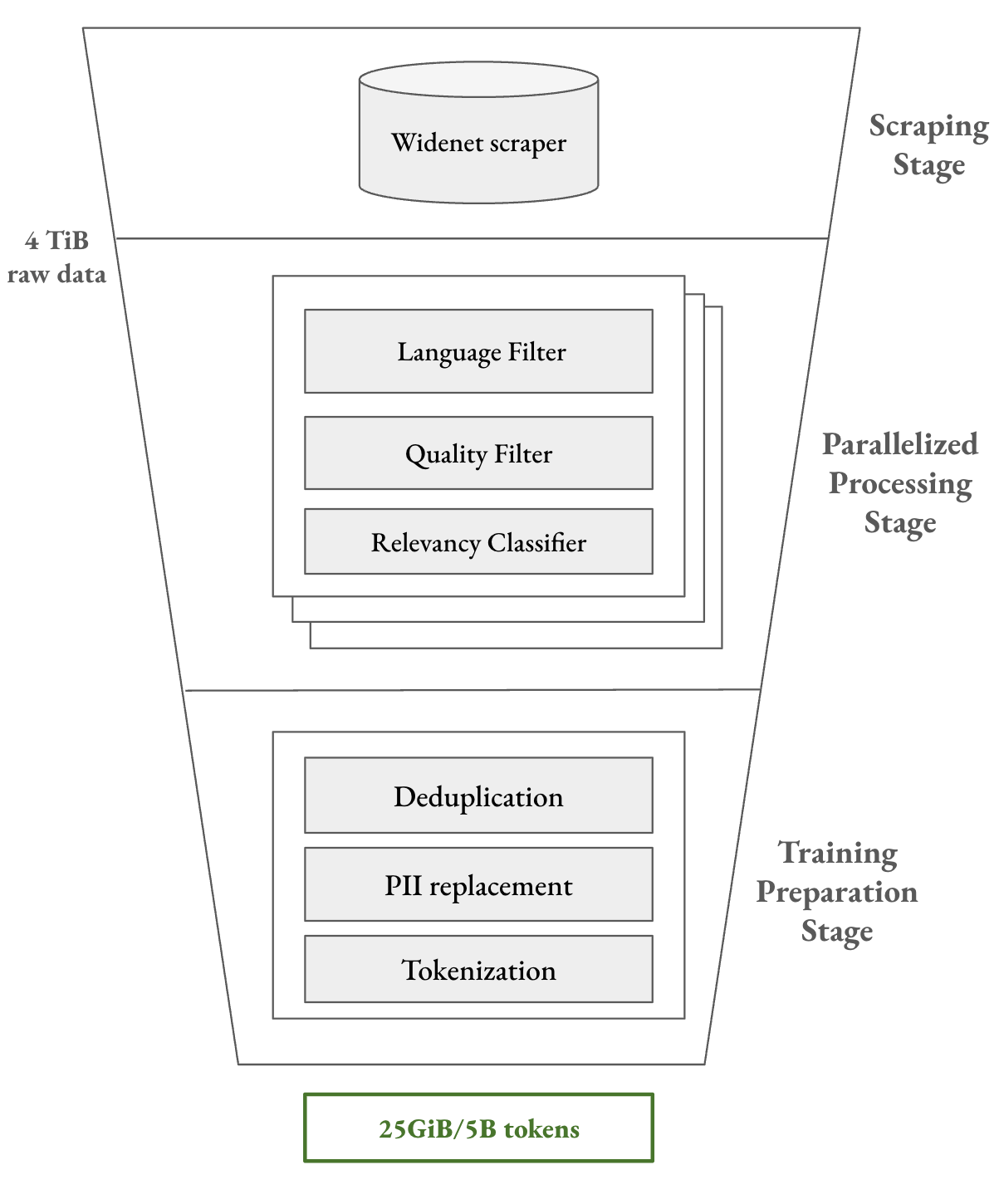}
\caption{\textbf{Three-stage data collection and processing pipeline:} (1) \textbf{Scraping Stage}—a wide-net scraper gathers 4 TiB of raw web content; (2) \textbf{Parallelized Processing Stage}—language filtering, quality filtering, and relevancy classification prune the data; (3) \textbf{Training Preparation Stage}—deduplication, PII replacement, and tokenization produce 25 GiB (about 5 billion tokens) of final training data.}

\label{fig:pipeline}
\end{figure}

We chose to build our dataset from scratch rather than filtering existing web-scale datasets like Hugging Face’s FineWeb \citep{fineweb}. Security-related content often appears ``non-English'' due to the presence of code, abbreviations, and fragmented sentences—leading to high perplexity. Datasets such as FineWeb \citep{fineweb} and The Pile \citep{the_pile} apply a definition of ``quality'' that does not align well with cybersecurity needs. For instance, FineWeb uses a fastText-based English filter with a threshold of \alert{0.65} \citep{fasttext}, which we found unsuitable for cybersecurity texts. This setting would exclude valuable content like CVE descriptions (see Appendix \ref{appendix:cve_description}). While existing datasets are valuable in their own contexts, we designed a focused data collection and curation pipeline—outlined in Figure~\ref{fig:pipeline}—tailored to the unique demands of cybersecurity.

\subsection{Data Collection via Wide-Net Scraper}\label{sec:tofu}

To expand our collection of cybersecurity data from sources that are harder to identify or navigate, we deployed a wide-net internet crawler to capture relevant content. The crawler begins by visiting a set of initial seed URLs, fetching and storing the raw page content, then parsing each response to extract new links, which are added to a processing queue. We seeded the crawler with URLs from a diverse set of cybersecurity domains.

The crawler prioritized URLs based on their depth in the search tree, favoring exploration within a domain before moving on to a new one. This produced a hybrid strategy—breadth-first search within a domain and depth-first search across domains.

To keep the crawl focused, we pruned irrelevant branches by applying a relevancy filter to the content of each scraped page. Links were only extracted if the page content passed this filter. Details of the relevancy filter are provided in Section \ref{sec:relevancy_filter}.

\subsection{Data Preprocessing}

While the top-of-funnel collection (Section \ref{sec:tofu}) gathered over \alert{4 TiB} of raw data, we built a scalable preprocessing pipeline to filter, clean, and transform the raw data into a high-quality, standardized dataset suitable for pretraining. We leveraged existing primitives to construct our own composable data pipelines, which were deployed across a cluster of virtual machines.

After running through this pipeline, the original \alert{4 TiB} of raw data was reduced to just under \alert{25 GiB} of cleaned text, yielding approximately \alert{0.6\%}. This low yield reflects both the extraction of only text content from noisy HTML pages and the removal of low-quality or irrelevant files. An analysis of the pipeline stages shows that about \alert{90\%} of the files were filtered out.

\subsubsection{Text Extraction} \label{sec:text_extraction}

Text extraction is the most computationally intensive and time-consuming part of the pipeline. This stage involved converting files of varying formats and sizes into clean Markdown files. We evaluated a range of closed and open-source tools on a sample dataset and developed a simple strategy to select the best tool for each file based on its format and size.

\subsubsection{Relevancy Filtering} \label{sec:relevancy_filter}

To ensure that our dataset consisted exclusively of cybersecurity-related documents, we implemented a relevance filtering pipeline. The initial filtering stage employed a keyword matching mechanism utilizing a curated list of approximately \alert{800} cybersecurity-related terms and acronyms. The presence of any keyword within a source document was treated as an indicator of relevance. While this approach offered computational efficiency and high recall, it suffered from a significant false positive rate.

To address the limitations of the keyword filter, we developed a small transformer-based classifier \citep{vaswani2017attention}. We constructed a labeled dataset comprising \alert{26,000} documents. Labels were generated using Gemini 2.0 Flash-Lite \citep{google_gemini_flashlite} prompted with an instruction to label each document as related/unrelated to cybersecurity. The resulting dataset was balanced, containing \alert{13,000} positive (relevant) and \alert{13,000} negative (irrelevant) samples, each under \alert{1024} tokens long.

\begin{wrapfigure}{r}{0.5 \textwidth}

\centering
\includegraphics[width=0.495\textwidth]{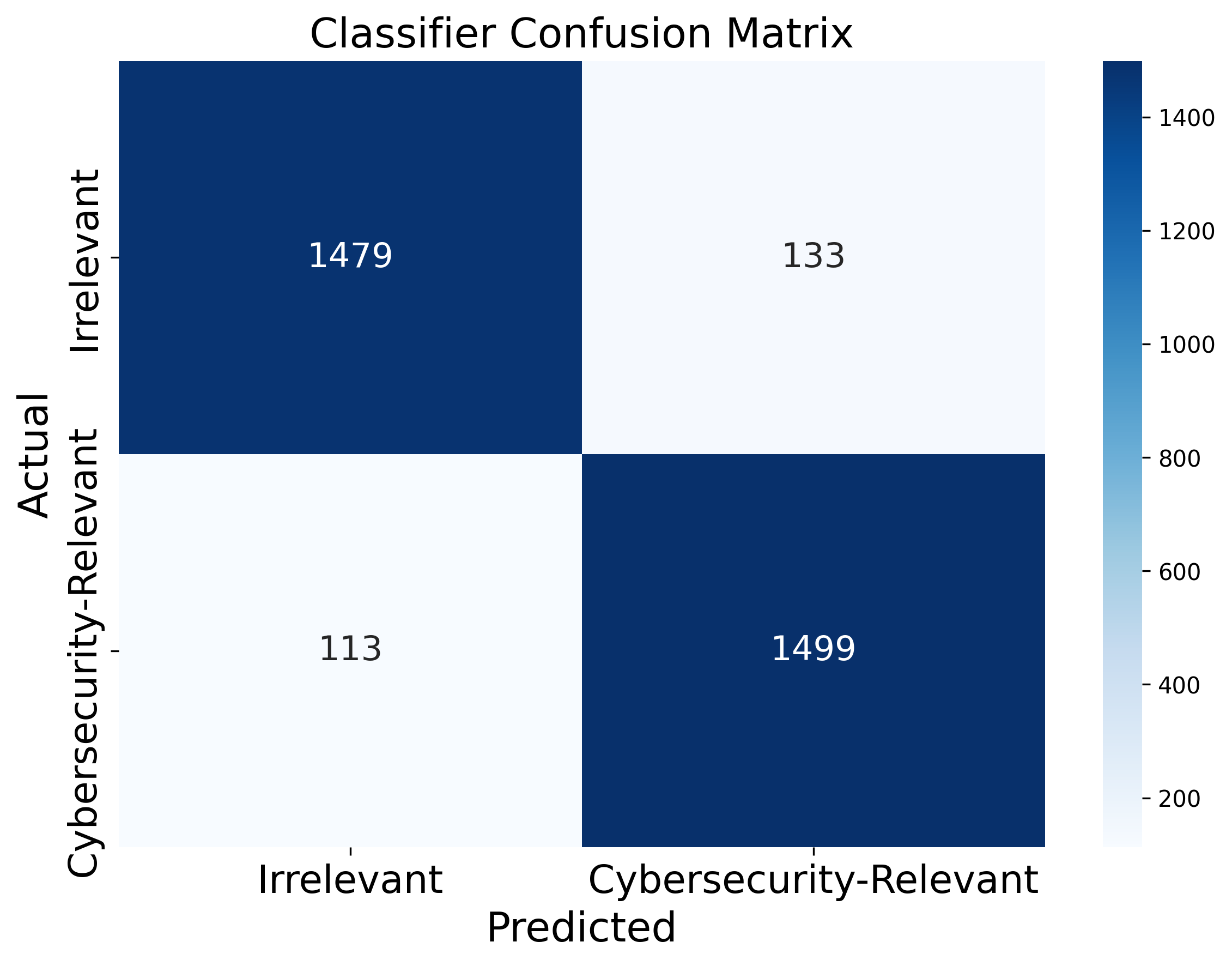}
\caption{Relevancy classifier on evaluation set}
\label{fig:classifier_confusion_matrix}
\end{wrapfigure}

We evaluated both the keyword-matching filter and the finetuned classifier on a held-out test set and found that the finetuned classifier significantly improved upon the keyword-matching filter, obtaining an F1 score of \alert{0.924} (see Figure \ref{fig:classifier_confusion_matrix}).

% \begin{figure}[ht!]

\subsubsection{Filter Evaluation Experiments}

In addition to our relevancy filter, we implemented additional filters to further remove low-quality data. In particular, we used a language filter and simple regex-based heuristics to identify files that were deficient or degenerate.

We conducted a series of experiments to assess the effectiveness of these filters within our data processing pipeline. In each experiment, we evaluated how well the filters retained high-quality, cybersecurity-relevant documents.

To carry out this analysis, we randomly sampled \alert{1,000} documents from our raw dataset. Each document was then assessed for quality and topic relevance using a frontier LLM, prompted to provide binary labels based on two criteria: (1) textual quality and (2) relevance to cybersecurity. A document was considered acceptable for inclusion only if it satisfied both criteria; otherwise, it should be filtered out.

We then applied the different filters independently to this labeled subset and compared their outputs against the ground truth. Some documents from key sources, including pages from the CVE and MITRE ATT\&CK databases, were excluded by these filters (see Appendix \ref{appendix:mitre_source}). Based on these evaluations, we ultimately chose to exclude most of the heuristic-based filters from our final data processing pipeline.

\subsection{Data Preparation}

Before training, we perform additional steps to clean and format the data for optimal use. To deduplicate our corpus, we applied an n-gram Bloom filter method \citep{bloom1970space} at the paragraph level. We also followed best practices to filter out personally identifiable information (PII) and prevent training on private data. Following prior work \citep{llama3, blakeney2024does, xie2023doremi, Gururangan2020-ks}, we upsampled known high-quality data related to Tactics, Techniques, and Procedures (TTPs). Finally, we split the resulting \alert{5.1 billion} tokens into a \alert{99\%} training set and a \alert{1\%} test set.

\section{Training and Evaluation}
\label{sec:evals}

We trained a \llamathree{}-8B model on the curated dataset through a further pretraining approach. Documents were packed into sequences of \alert{4096} tokens for maximum efficiency \citep{t5}. Training was performed using \deepspeed{} \citep{deepspeed} on a multi-node compute cluster. We used the AdamW optimizer \citep{loshchilov2017decoupled} with a cosine decay learning rate schedule \citep{loshchilov2017sgdr}.

% \subsection{Evaluation}

\paragraph{Benchmarks}
We consider the benchmarks detailed in Section \ref{sec:benchmarks} in our experiments. In Appendix \ref{app:additional_benchmarks}, we also discuss other cybersecurity benchmarks from the literature that we reviewed but chose not to include in our evaluation, as they were misaligned with our evaluation objectives.

\begin{table}[thpb]
  \centering
  \resizebox{\textwidth}{!}{%
    \begin{threeparttable}
    \begin{tabular}{l cccc}
      \toprule
      \textbf{Model}
        & \textbf{\ctibench{}-MCQA}
        & \textbf{\ctibench{}-RCM}
        & \textbf{CyberMetric-500}
        & \textbf{SecBench} \\
      \midrule
      \securityllm
        & 0.601$\pm$0.009
        % & 0.218$\pm$0.96\tnote{*}
        & N/A$^{\dagger}$
        & 0.807$\pm$0.010
        & 0.668$\pm$0.012 \\
      \llamathree-8B
        & 0.623$\pm$0.012
        & 0.630$\pm$0.024
        & 0.848$\pm$0.008
        & 0.735$\pm$0.011 \\
      \whiterabbit-8B
        & 0.629$\pm$0.011
        & 0.631$\pm$0.031
        & 0.846$\pm$0.006
        & 0.738$\pm$0.010 \\
      \primus-Seed-8B
        & 0.660$\pm$0.009
        & 0.622$\pm$0.014
        & 0.853$\pm$0.007
        & 0.732$\pm$0.009 \\
      \midrule
      \llamathree-70B
        & \textbf{0.678$\pm$0.008}
        & 0.709$\pm$0.010
        & \textbf{0.918$\pm$0.008}
        & 0.832$\pm$0.008 \\
      \whiterabbit-70B
        & \textbf{0.680$\pm$0.007}
        & 0.711$\pm$0.008
        & \textbf{0.917$\pm$0.009}
        & \textbf{0.836$\pm$0.008} \\
      \midrule
      \gptfourOmini
        & 0.670$\pm$0.003
        & 0.659$\pm$0.002
        & 0.889$\pm$0.002
        & 0.800$\pm$0.002 \\
      \secgemini
        & 0.8630
        & 0.8610
        & N/A$^\ddagger$
        & N/A$^\ddagger$ \\
      \midrule
      \textbf{\ModelName{}-8B}
        & 0.662$\pm$0.007 ($\uparrow$6.26\%)
        & \textbf{0.720$\pm$0.017} ($\uparrow$14.29\%)
        & 0.848$\pm$0.009 ($\leftrightarrow$0.0\%)
        & 0.723$\pm$0.009 ($\downarrow$1.63\%) \\
      \bottomrule
    \end{tabular}
    \begin{tablenotes}
      \item[$\dagger$]Due to its limited context length, \securityllm{} did not produce meaningful results on \ctibench{}-RCM.
      \item[$\ddagger$]Results are not reported by the developer.
    \end{tablenotes}
    \end{threeparttable}
  }
    \vspace{1ex}
    \caption{Performance of the models on the selected cybersecurity benchmarks (temperature 0.3). Reported performance differences are relative to the \llamathree{}-8B model, which \ModelName{} is based on.}
    \label{tab:llm-cyber}
\end{table}

\begin{table}[thpb]
  \centering
  \resizebox{\textwidth}{!}{%
    \begin{threeparttable}
    \begin{tabular}{l cccc}
      \toprule
      \textbf{Model}
        & \textbf{\ctibench{}-MCQA}
        & \textbf{\ctibench{}-RCM}
        & \textbf{CyberMetric-500}
        & \textbf{SecBench} \\
      \midrule
      \llamathree-8B
        & 0.641 
        & 0.636
        & 0.849
        & 0.744 \\
      \llamathree-70B
        & 0.689 
        & 0.710
        & 0.924 
        & 0.839 \\
      \midrule
      \textbf{\ModelName{}-8B}
        & 0.676 ($\uparrow$5.46\%)
        & 0.724 ($\uparrow$13.84\%)
        & 0.851 ($\uparrow$0.24\%)
        & 0.734 ($\downarrow$1.34\%)\\ 
      \bottomrule
    \end{tabular}
    \end{threeparttable}
  }
    \vspace{1ex}
    \caption{Performance of the models on the selected cybersecurity benchmarks (temperature 0). Reported performance differences are relative to the \llamathree{}-8B model, which \ModelName{} is based on.}
    \label{tab:llm-cyber-zero}
\end{table}

\paragraph{Baselines}
The models reviewed in Section \ref{sec:llms_for_security} are included as baselines for comparison. For further details, we refer readers to their respective model cards on Hugging Face.

\paragraph{Input Formatting and Prompting}
To ensure fair evaluation, we carefully design the prompting strategy, as poor formatting can obscure a model’s knowledge and capability. 
Since log-probabilities are unavailable for closed models, we evaluate based on predicted tokens. 
Pretrained models, which do not follow instructions directly, are evaluated using 5-shot prompts to help infer both the task and the expected output format. 
Instruct-finetuned (IFT) models, on the other hand, are evaluated in a zero-shot setting to avoid inconsistent or verbose completions. 
We prompt them only with the question and extract their answers using regex (regular expressions). 
Additional implementation details, including templates and regex patterns, are provided in Appendices \ref{app:prompt_design} and \ref{app:regex}.

\section{Results}
Evaluations were conducted over \alert{10} trials to account for stochasticity in token sampling (with temperature set to \alert{0.3}) and in the selection of few-shot examples. 
We report mean accuracy with the standard deviation over 10 trials. 
Results are shown in Table \ref{tab:llm-cyber}. 
We also evaluate the models at temperature 0 and present those results in Table \ref{tab:llm-cyber-zero}.

\subsection{Security Benchmark Performance}

\ModelName-8B achieves \textbf{state-of-the-art performance} for its size class on the selected CTI tasks, performing \textbf{on par with \gptfourOmini{}}. It trails \gptfourOmini{} by only 0.8 points on \ctibench{}-MCQA while outperforming it by 6.1 points on \ctibench{}-RCM. \ModelName-8B also surpasses both \llamathree-70B and \whiterabbit-70B by about 1 point on \ctibench{}-RCM, while falling short by less than 2 points on \ctibench{}-MCQA.

The model achieves a notable improvement of over 3 points in accuracy compared to its parent model, \llamathree-8B-base. We emphasize that, since we evaluate general knowledge rather than a task-specific fine-tuning setup, these gains are substantially large for improvements over a general-purpose model on a specialized, nuanced domain of knowledge—a trend also observed in prior work \citep{meditron, whiterabbitneo, primus}.

Overall, our results position \ModelName-8B as the leading 8-billion parameter model for CTI security tasks.

\subsection{General Benchmark Performance}
Prior work \citep{sun2024dial, chen2025continual} shows that continued pretraining is typically followed by evaluations on general benchmarks to check for catastrophic forgetting. Following this practice, we verify that our training does not significantly degrade general performance. 
We evaluate our model on the full MMLU benchmark. \ModelName-8B achieves a score of \alert{0.593$\pm$0.004}, compared to \alert{0.617$\pm$0.004} for \llamathree-8B. The observed \alert{2.4} point drop is consistent with prior literature.

\section{Use Cases}
\label{sec:use_cases}
Large language models in cybersecurity are no longer just theoretical—they are increasingly being integrated into high-value, operational workflows across the security lifecycle. \ModelName{} was developed with these real-world applications in mind. We outline three core areas where \ModelName{} is \textbf{currently being piloted or deployed}, either in its pretrained form or after fine-tuning.

\paragraph{SOC Acceleration}

Security Operations Centers (SOCs) face a constant stream of alerts that require triage, enrichment, and contextualization. \ModelName{} is currently being piloted to automate key parts of this workflow. When finetuned or prompted effectively, it is used to:
\begin{itemize}
    \item Summarize multi-source alerts into human-readable case notes
    \item Generate incident timelines and identify relevant entities
    \item Draft analyst-style reports to support incident resolution and handoffs
\end{itemize}

These capabilities reduce time-to-triage and enable analysts to handle more alerts with higher accuracy.

\paragraph{Proactive Threat Defense}

Beyond reactive workflows, \ModelName{} is being deployed to model and simulate attacker behavior. This includes:
\begin{itemize}
    \item Extracting Tactics, Techniques, and Procedures (TTPs) from threat intelligence reports
    \item Prioritizing vulnerabilities based on contextual impact and exploitability
    \item Generating attack path hypotheses from asset and configuration data
    \item Drafting penetration test reports with vulnerability details and remediation steps
\end{itemize}

A particularly effective application has been fine-tuning \ModelName{} for MITRE ATT\&CK Technique extraction from unstructured threat reports. In internal evaluations, \ModelName{} outperformed a similarly sized non-security-tuned model (\llamathree{}-8B) by over 10\% on this classification task, highlighting the value of security-domain pretraining.

\paragraph{Engineering Enablement}

Security engineering and platform teams often face the challenge of enforcing security standards across rapidly evolving development environments. \ModelName{} is being used to streamline and enhance these workflows by providing secure development guidance, validating configurations, and supporting compliance efforts.

When applied effectively, \ModelName{} helps teams:
\begin{itemize}
    \item Interpret and apply security policies during development and deployment
    \item Validate configuration files and infrastructure setups against best practices
    \item Assess whether submitted evidence meets compliance control requirements
    \item Analyze security policies for inconsistencies and outdated controls
\end{itemize}

These tasks typically require a combination of context awareness and domain-specific knowledge—areas where \ModelName{} excels.

\noindent If you are interested in using \ModelName{} for your own cybersecurity applications or research, please reach out to \textbf{Paul Kassianik} (\href{mailto:paulkass@cisco.com}{\texttt{paulkass@cisco.com}}) or \textbf{Dhruv Kedia} (\href{mailto:dkedia@cisco.com}{\texttt{dkedia@cisco.com}}) at Foundation AI.

\section{Conclusion and Future Work}

In this work, we introduced \ModelName, a cybersecurity-specialized large language model built upon \llamathree{}. Addressing key limitations that have hindered LLM adoption in cybersecurity, we curated a high-quality cybersecurity dataset and demonstrated significant improvements in task-specific performance. Our evaluations show that \ModelName{} achieves capabilities competitive with much larger models, such as \llamathree{}-70B, without compromising general-purpose functionality.

We highlight several promising directions for further research and development:
\begin{itemize}
    \item Scale up \ModelName{} by increasing its parameter count and expanding the training corpus.
    \item Extend \ModelName{} to handle cybersecurity-related coding tasks.
    \item Integrate \ModelName{} into tool-calling and agentic systems for more interactive applications.
\end{itemize}

By releasing \ModelName{} publicly, we aim to support both the cybersecurity and AI research communities, fostering broader experimentation, advancement, and practical deployment of AI-driven security tools. We envision \ModelName{} accelerating LLM adoption among cybersecurity professionals and enabling new lines of security research.

We also hope this work inspires future studies on optimizing specialized pretraining techniques, expanding coverage of cybersecurity domains, and investigating how smaller models can rival or surpass larger general-purpose LLMs in domain-specific tasks. Ultimately, \ModelName{} highlights the impact of targeted, domain-aware training in making LLMs more effective for secure, intelligent cybersecurity operations.

\section*{Acknowledgements}

We are grateful to Zane Mogannam, Divit Rawal, Erica Liu, Arjun Banerjee, Neel Kolhe, Alena Subach, and Kathryn Sun from UC Berkeley Launchpad for their help with data collection and processing.

We thank Sasha Sinkevich and Jon McLachlan from YSecurity for their early efforts in kick-starting this project.

We also sincerely thank Cisco’s Security \& Trust Organization (S\&TO) for their ongoing partnership in identifying and validating impactful use cases for this model. In particular, we acknowledge Omar Santos, Robert Kerby, Vinay Bansal, Aaron Carter, Chalamaiah Gupta Reddy, and Sam Cosentino, whose early input shaped our understanding of real-world security workflows and informed our development strategy.

Finally, we thank Kamile Lukosiute for her valuable feedback during the preparation of this report.

% We are grateful for Crusoe AI\footnote{\url{https://www.crusoe.ai/}} for providing us with hardware to train these models.

\bibliography{bibliography}

\newpage

\appendix

\section{Data Processing Details}\label{appendix:data_details}
% \subsection{Relevancy Filter Keyword Matching}\label{appendix:keyword_matching}
% The below is the list of cybersecurity-related keywords used as part of the keyword-matching relevancy filter.

% \begin{framed}
% \small
% % \ttfamily
% \input{cybersecurity_keywords.txt} 
% \end{framed}

% \subsection{Labelling Instruction Prompt}\label{appendix:labelling_prompt}
% Below is the instruction prompt used for labeling the training data used in ModernBERT classifier. 
% \begin{framed}
% Analyze the following webpage content and classify it based on two criteria: 

% 1. Is the content low quality for foundation model training? Consider factors like redundancy, spam, scraped content, excessive advertisements, or poor formatting.

% 2. Is the content related to cybersecurity? Look for discussions on malware, hacking, security threats, vulnerabilities, or cybersecurity best practices. 

% Return a JSON response with two keys: low\_quality (true/false) and cybersecurity\_related (true/false).

% Content:\{doc\_content\}
% \end{framed}

\subsection{Example of CVE Description}\label{appendix:cve_description}
Below is an example of a CVE description that scores below 0.65 on the fastText language filter \citep{fasttext} but is considered a high-quality cybersecurity document.

\begin{mdframed}[everyline=true]
\begin{lstlisting}[
    numbers=none,
    frame=none,
    framesep=5pt,
    xleftmargin=5pt,
    xrightmargin=5pt,
    basicstyle=\normalfont\small,
    breaklines=true,
    breakautoindent=false,
    breakindent=0pt,
    showstringspaces=false,
    columns=fullflexible
]
Advisory ID: CVE-2016-6335

Summary: No summary provided. 

Details:
MediaWiki before 1.23.15, 1.26.x before 1.26.4, and 1.27.x before 1.27.1 does not generate head items in the context of a given title, which allows remote attackers to obtain sensitive information via a parse action to api.php.

Published: 2017-04-20T17:59:00Z
Modified: 2024-09-18T02:36:06.797830Z

Affected Packages:
- mediawiki (Debian:11)
  Introduced in: 0
  Fixed in: 1:1.27.1-1
  Source: https://storage.googleapis.com/cve-osv-conversion/osv-output/CVE-2016-6335.json
- mediawiki (Debian:12)
  Introduced in: 0
  Fixed in: 1:1.27.1-1
  Source: https://storage.googleapis.com/cve-osv-conversion/osv-output/CVE-2016-6335.json
- mediawiki (Debian:13)
  Introduced in: 0
  Fixed in: 1:1.27.1-1
  Source: https://storage.googleapis.com/cve-osv-conversion/osv-output/CVE-2016-6335.json
- Unknown package (Unknown ecosystem)
  Introduced in: 0
  Source: https://storage.googleapis.com/cve-osv-conversion/osv-output/CVE-2016-6335.json
 References:
- https://lists.wikimedia.org/pipermail/mediawiki-announce/2016-August/000195.html
- https://phabricator.wikimedia.org/T139570
- https://bugzilla.redhat.com/show_bug.cgi?id=1369613
- https://phabricator.wikimedia.org/T139565
- https://security-tracker.debian.org/tracker/CVE-2016-6335
\end{lstlisting}
\end{mdframed}

\subsection{Example of a MITRE ATT\&CK Page}\label{appendix:mitre_source}
Below is an example of a page from the MITRE ATT\&CK database that fails the quality filter of a well-known text filtering system but is considered a high-quality cybersecurity document.

\begin{mdframed}[everyline=true]
\begin{lstlisting}[
    numbers=none,
    frame=none,
    framesep=5pt,
    xleftmargin=5pt,
    xrightmargin=5pt,
    basicstyle=\normalfont\small,
    breaklines=true,
    breakautoindent=false,
    breakindent=0pt,
    showstringspaces=false,
    columns=fullflexible
]
Description: Adversaries may gather credentials from the proc filesystem or `/proc`. The proc filesystem is a pseudo-filesystem used as an interface to kernel data structures for Linux based systems managing virtual memory. For each process, the `/proc/<PID>/maps` file shows how memory is mapped within the process's virtual address space. And `/proc/<PID>/mem`, exposed for debugging purposes, provides access to the process's virtual address space.Huseyin Can YUCEEL & Picus Labs. (2022, March 22).baeldung. (2022, April 8). Understanding the Linux /proc/id/maps File. When executing with root privileges, adversaries can search these memory locations for all processes on a system that contain patterns indicative of credentials. Adversaries may use regex patterns, such as <code>grep -E "^[0-9a-f-]* r" /proc/"\$pid"/maps | cut -d' ' -f 1</code>, to look for fixed strings in memory structures or cached hashes.Atomic Red Team. (2023, November). T1003.007 - OS Credential Dumping: Proc Filesystem. When running without privileged access, processes can still view their own virtual memory locations. Some services or programs may save credentials in clear text inside the process's memory.Gregal, H. (2017, May 12). MimiPenguin.Carlos Polop. (2023, March 5). Linux Privilege Escalation. If running as or with the permissions of a web browser, a process can search the `/maps` & `/mem` locations for common website credential patterns (that can also be used to find adjacent memory within the same structure) in which hashes or cleartext credentials may be located.

Domain: Enterprise Attack

Tactics: Credential Access

Detection: To obtain the passwords and hashes stored in memory, processes must open a maps file in the `/proc` filesystem for the process being analyzed. This file is stored under the path `/proc/PID/maps`, where the `PID` directory is the unique pid of the program being interrogated for such authentication data. The AuditD monitoring tool, which ships stock in many Linux distributions, can be used to watch for hostile processes opening this file in the proc file system, alerting on the pid, process name, and arguments of such programs.

Platforms: Linux

Data Sources: Command: Command Execution, File: File Access

Sub-Technique Of: T1003
\end{lstlisting}
\end{mdframed}

\section{Evaluation Details}
\label{app:eval_details}

\subsection{Implementation}
We used the vLLM framework \citep{vllm}, an open-source, high-throughput engine optimized for inference. vLLM introduces paged attention, a memory management technique that decouples attention computation from memory layout, enabling efficient dynamic batching and reducing memory fragmentation. This design improves GPU utilization by supporting continuous-token streaming and lowering overhead in multi-query processing. The system also supports asynchronous execution and tensor parallelism, allowing it to scale across multiple GPUs. Compatible with Hugging Face, vLLM offers an optimized runtime with low latency, making it well-suited for production-grade LLM serving. Additional details are available on their website\footnote{\url{https://docs.vllm.ai/en/stable/serving/offline_inference.html}}.

\subsection{Additional Benchmarks}
\label{app:additional_benchmarks}

The following benchmarks, although recognized in the literature, were excluded from our evaluations as they were either out of scope for our study or could have led to misleading conclusions. The specific justifications are detailed below.

\paragraph{MMLU -- computer security}
Although highly relevant, we did not report results on the computer security section of MMLU as it contains only 100 samples, which could lead to statistically unreliable conclusions.

\paragraph{CyberSecEval-3}
The CyberSecEval benchmarks \citep{cyberseceval}, part of the Purple Llama suite \citep{purple_llama}, form a comprehensive collection of tasks aimed at evaluating the cybersecurity vulnerabilities of LLMs. CyberSecEval-3 assesses eight types of risks across two broad categories: risks to third parties, and risks to application developers and end users. While we recognize the significance of this suite, we excluded it from our evaluation because it focuses on model robustness—such as resistance to prompt injections, spear phishing, and autonomous offensive cyber operations—rather than assessing cybersecurity knowledge.

\paragraph{SecEval}
SecEval \citep{li2023seceval} provides over 1200 MCQA questions (English section) spanning nine cybersecurity domains. 
The dataset is generated by prompting GPT-4 with content from authoritative sources, including open-licensed textbooks, official documentation, and industry standards. 
We excluded SecEval from our evaluation as it appears highly \textit{saturated}—many models, regardless of size, architecture, or pretraining corpus (e.g., \llamathree{}-8B achieves ~86\% while \gptfour{} achieves ~90\%), exhibit similar performance levels.
 
\paragraph{SECURE}
\textbf{SEC}urity Extraction, \textbf{U}nderstanding \& \textbf{R}easoning \textbf{E}valuation \citep{secure} was developed to assess model performance in realistic cybersecurity scenarios. It includes six datasets focused on the Industrial Control System (ICS) domain, evaluating knowledge extraction, comprehension, and reasoning using industry-standard sources. We also omit SECURE due to its saturation; for instance, \llamathree{}-8B scores ~83\% and \gptfourO{} scores ~90\%.

\paragraph{SecQA}
The MCQA dataset SecQA was generated by \gptfour{} using content from the textbook \textit{Computer Systems Security: Planning for Success} \citep{computer_sys_security_book}. 
Although it is a recognized resource in the field, we excluded it from our evaluation due to its small size (around 120 samples).

\subsection{Prompt Design}
\label{app:prompt_design}

Designing an effective prompting strategy is essential to fairly evaluate a model’s knowledge, as improper phrasing or formatting can obscure what the model actually knows and lead to misleading conclusions. \alert{Since log-probabilities are not accessible in closed models, we evaluate based on the model’s predicted next tokens.}

\subsubsection{Pretrained Models}
Base pretrained LMs do not follow instructions explicitly but instead complete the given input text. Therefore, we leverage their few-shot learning capabilities. We use 5-shot prompts to help the model infer both the task and the expected answer format (e.g., MCQA or CWE ID mapping).

When available, we construct these examples from the benchmark’s development (dev) set. Among our selected benchmarks, only MMLU includes a dev set; for the others, we sample examples directly from the dataset. Since we run multiple trials and report the average performance, the effect of this stochasticity is also averaged out. In each trial, the model sees a different set of examples, ensuring prompt diversity and randomness.

Lastly, although pretrained models are not instruction-following by design, we append a brief sentence—``\texttt{The following are multiple choice questions about computer security.}''—to help them infer the task context. This is a common practice in evaluations, as seen in benchmarks for models like \llamathree{}.

The input templates for pretrained models are provided for both MCQA and CWE ID mapping (i.e., \ctibench{}-RCM) in Figures \ref{fig:pretrained_input_tmp_mcqa} and \ref{fig:pretrained_input_tmp_cwe}.

\begin{figure}[t]
\centering
\begin{lstlisting}[
    frame=single,
    framesep=5pt,
    xleftmargin=5pt,
    xrightmargin=5pt,
    basicstyle=\small\ttfamily\linespread{1.2}\selectfont,
    breaklines=true,
    breakautoindent=false,
    breakindent=0pt,
    numbers=none
]
The following are multiple choice questions about computer security.

The ____________ is anything which your search engine cannot search.
A. Haunted web
B. World Wide Web
C. Surface web
D. Deep Web
Answer: D

Exploitation of the Heartbleed bug permits
A. overwriting cryptographic keys in memory
B. a kind of code injection
C. a read outside bounds of a buffer
D. a format string attack
Answer: C

.
.
.

Three of the following are classic security properties; which one is not?
A. Confidentiality
B. Availability
C. Correctness
D. Integrity
\end{lstlisting}
\caption{Few-shot prompt format used with pretrained models for MCQA tasks. The model is expected to respond in the format ``\texttt{Answer: X},'' where X is one of A, B, C, or D.}
\label{fig:pretrained_input_tmp_mcqa}
\end{figure}

\begin{figure}[t]
\centering
\begin{lstlisting}[
    frame=single,
    framesep=5pt,
    xleftmargin=5pt,
    xrightmargin=5pt,
    basicstyle=\small\ttfamily\linespread{1.2}\selectfont,
    breaklines=true,
    breakautoindent=false,
    breakindent=0pt,
    numbers=none
]
The following is a CVE description. Map it to the appropriate CWE ID.

CVE Description: A SQL injection vulnerability exists in Novel-Plus v4.3.0-RC1 and prior. An attacker can pass specially crafted offset, limit, and sort parameters to perform SQL injection via /novel/userFeedback/list.
Answer: CWE-89

CVE Description: Cross-Site Request Forgery (CSRF) vulnerability in Doofinder WP & WooCommerce Search.This issue affects Doofinder WP & WooCommerce Search: from n/a through 2.0.33.
Answer: CWE-352

.
.
.

CVE Description: Tenda AX1803 v1.0.0.1 contains a stack overflow via the iptv.city.vlan parameter in the function getIptvInfo.
\end{lstlisting}
\caption{Few-shot prompt format used with pretrained models for the CWE ID mapping task (\ctibench{}-RCM). The model is expected to respond in the format ``\texttt{Answer: CWE-X},'' where X is a unique CWE ID.}
\label{fig:pretrained_input_tmp_cwe}
\end{figure}

\subsubsection{Instruction-Finetuned Models}
Instruct-finetuned models are designed to follow specific prompt formats but often produce inconsistent outputs in few-shot settings. Instead of returning a clean ``\texttt{Answer: }'' format, they may prepend phrases like ``Sure, here’s the answer...,'' which breaks format consistency and complicates evaluation.

To mitigate this, we evaluate IFT models in a zero-shot setting—prompting only with the question and extracting the answer using regex. That is, no examples are included in the input. To guide the model’s output format, we append a short instruction of 1–2 sentences. If the benchmark provides its own instruction, we use it directly. Otherwise, we adopt the fusion of the instructions used in \llamathree{} evaluations (based on MMLU) and OpenAI Simple Evals\footnote{\url{https://github.com/openai/simple-evals}}: \alert{``\texttt{Given the following question and four candidate answers (A, B, C and D), choose the best answer. Your response should be of the following format: 'Answer: \$LETTER' (without quotes) where LETTER is one of A, B, C, or D.}''} Among our benchmarks, only \ctibench{} provides its own instruction. No system prompts are used in any evaluation.

Example input prompts for chat models in MCQA and CWE ID mapping tasks are shown in Figures \ref{fig:ift_input_tmp_mcqa} and \ref{fig:ift_input_tmp_cwe}.

\begin{figure}[t]
\centering
\begin{lstlisting}[
    frame=single,
    framesep=5pt,
    xleftmargin=5pt,
    xrightmargin=5pt,
    basicstyle=\small\ttfamily\linespread{1.2}\selectfont,
    breaklines=true,
    breakautoindent=false,
    breakindent=0pt,
    numbers=none
]
Given the following question and four candidate answers (A, B, C and D), choose the best answer. Your response should be of the following format: 'Answer: $LETTER' (without quotes) where LETTER is one of A, B, C, or D.

Three of the following are classic security properties; which one is not?
A. Confidentiality
B. Availability
C. Correctness
D. Integrity
\end{lstlisting}
\caption{Zero-shot prompt format used with instruct-finetuned models for MCQA tasks.}
\label{fig:ift_input_tmp_mcqa}
\end{figure}

\begin{figure}[t]
\centering
\begin{lstlisting}[
    frame=single,
    framesep=5pt,
    xleftmargin=5pt,
    xrightmargin=5pt,
    basicstyle=\small\ttfamily\linespread{1.2}\selectfont,
    breaklines=true,
    breakautoindent=false,
    breakindent=0pt,
    numbers=none
]
Analyze the following CVE description and map it to the appropriate CWE. Provide a brief justification for your choice. Ensure the last line of your response contains only the CWE ID.

CVE Description: Tenda AX1803 v1.0.0.1 contains a stack overflow via the iptv.city.vlan parameter in the function getIptvInfo.
\end{lstlisting}
\caption{Zero-shot prompt format used with instruct-finetuned models for the CWE ID mapping task (\ctibench{}-RCM).}
\label{fig:ift_input_tmp_cwe}
\end{figure}

\subsection{Postprocessing and Answer Extraction}
\label{app:regex}

We apply several regular expression (regex) operations to the model’s output string responses.

\subsubsection{MCQA Tasks}

\paragraph{Correct Format}
We start by matching a case-insensitive ``\texttt{answer:}''—allowing optional spaces and an optional opening parenthesis—followed by a letter A–D, and optionally ending with a closing parenthesis or a word boundary. This captures variations such as ``\texttt{Answer: C}'' or ``\texttt{answer: (B)}''. If no match is found in this format, we flag the sample as ``misformatted'' and proceed with the ``misformatted patterns'' described below. Note that pretrained models almost never misformat their responses, so these operations are primarily applied to chat models.

\begin{enumerate}
    \item \textbf{Responses in ``\texttt{answer is}'':} Matches case-insensitive patterns like ``\texttt{Answer is A}'' or ``\texttt{answer is: (D)}'', allowing an optional colon, optional ``('', and an uppercase A–Z, optionally ending with ``)'' or a word boundary.
    \item \textbf{Single Letter Responses:} Captures standalone uppercase letters A–Z, optionally enclosed in parentheses, such as ``\texttt{C}'' or ``\texttt{(B)}''.
    \item \textbf{Use of ``\texttt{option}'' instead of ``\texttt{answer}'':} Matches case-insensitive ``option'' followed by an uppercase A–Z at a word boundary, e.g., ``\texttt{Option A}''.
\end{enumerate}

\subsubsection{CWE ID Mapping Task}
Since this task is part of \ctibench{}, we use the authors' codebase\footnote{\url{https://github.com/xashru/cti-bench}} to extract CWE IDs. The extraction process is similar to that of MCQA answers, where we search for the pattern ``\texttt{CWE-X},'' with X representing a sequence of digits.

\end{document}

%% file: packages.tex
\usepackage{graphicx} % Required for inserting images
\usepackage[preprint,nonatbib]{neurips_2025}  % Add numberbib option
\usepackage{array, booktabs, tabularx}
\usepackage[numbers, sort]{natbib}  % Let the NeurIPS package handle natbib
\usepackage{times,mathptmx}
\usepackage[noblocks]{authblk}
\usepackage[utf8]{inputenc}
\usepackage{fontenc}
\usepackage[hyphens]{url}
\usepackage{hyperref}
\hypersetup{hidelinks=true, colorlinks=true, citecolor=blue, linkcolor=black, urlcolor=black}
\usepackage[textsize=scriptsize]{todonotes}
\usepackage{amsmath}
\usepackage{algorithm}             % float environment
\usepackage{algpseudocode}         % \begin{algorithmic} …

\usepackage{listings}
\usepackage{xcolor}

\usepackage{needspace}  % in your preamble
\Needspace{8\baselineskip}

\usepackage{makecell}
\usepackage{inconsolata}

\usepackage{fvextra}
\usepackage{caption}
\usepackage{pgfplots}
\pgfplotsset{compat=1.18}
\usepackage{pgf-pie}
\usepackage{fancyvrb}
\usepackage{mdframed}
\usepackage{arydshln}
\usepackage{threeparttable}
\usepackage{subcaption}
\usepackage{wrapfig}